\documentclass{statsoc}\usepackage{fix-cm}

\usepackage[a4paper]{geometry}
\usepackage{graphicx}
\usepackage[textwidth=8em,textsize=small]{todonotes}
\usepackage{amsmath}
\usepackage{amsfonts}
\usepackage{natbib}
\usepackage{booktabs}
\usepackage{etoolbox}
\usepackage{url}
\usepackage{enumitem}
\newcommand{\Var}{\text{Var}}

\usepackage[caption = false]{subfig} 

\makeatletter
\patchcmd{\@makecaption}
  {\parbox}
  {\advance\@tempdima-\fontdimen2} 
  {}{}
\makeatother

\newtheorem{theorem}{Theorem}

\title[Surrogate Primary Correlation]{Bias in Meta-Analytic Modeling of Surrogate Endpoints in Cancer Screening Trials}
\author[James P. Long]{James P. Long}
\address{University of Texas MD Anderson Cancer Center,
Houston,
Texas.}

\email{jplong@mdanderson.org}

\author{Abhishikta Roy}
\address{University of Texas School of Public Health,
Houston,
Texas.}

\author{Ehsan Irajizad}
\address{University of Texas MD Anderson Cancer Center,
Houston,
Texas.}

\author{Kim-Anh Do}
\address{University of Texas MD Anderson Cancer Center,
Houston,
Texas.}

\author{Yu Shen}
\address{University of Texas MD Anderson Cancer Center,
Houston,
Texas. }

\email{yshen@mdanderson.org}

\begin{document}
\begin{abstract}

In meta-analytic modeling, the functional relationship between a primary and surrogate endpoint is estimated using summary data from a set of completed clinical trials. Parameters in the meta-analytic model are used to assess the quality of the proposed surrogate. Recently, meta-analytic models have been employed to evaluate whether late-stage cancer incidence can serve as a surrogate for cancer mortality in cancer screening trials. A major challenge in meta-analytic models is that uncertainty of trial-level estimates affects the evaluation of surrogacy, since each trial provides only estimates of the primary and surrogate endpoints rather than their true parameter values. In this work, we show via simulation and theory that trial-level estimate uncertainty may bias the results of meta-analytic models towards positive findings of the quality of the surrogate. We focus on cancer screening trials and the late stage incidence surrogate. We reassess correlations between primary and surrogate endpoints in Ovarian cancer screening trials. Our findings indicate that completed trials provide limited information regarding quality of the late-stage incidence surrogate. These results support restricting meta-analytic regression usage to settings where trial-level estimate uncertainty is incorporated into the model.

\end{abstract}

\section{Introduction}

Cancer screening aims to reduce mortality by detecting tumors at an early stage, when treatments are more likely to be effective. 
Screening programs have been widely used in clinical practice including mammograms for breast cancer, sigmoidoscopy for colorectal cancer, and low-dose computed tomography (LDCT) scans for lung cancer in smokers \citep{uspstf2024}. New imaging and molecular diagnostic technologies have led to the development of promising new screening techniques, which are at various stages of scientific and commercial development \citep{klein2021clinical,irajizad2023mortality,csrn2024}. 

Before being widely adopted, a screening method should be tested in clinical trials to demonstrate benefit to people at average or increased risk of the cancer. These trials typically randomize subjects to two arms: a standard of care (control) arm and a screening (intervention) arm. Cancer-specific mortality is often considered the gold-standard primary endpoint in screening trials \citep{minasian2023study,malagon2024time}. 

Unfortunately, using reduction in cancer-specific mortality to demonstrate efficacy is costly, both in length of follow up time and trial sample size. For example, the National Lung Screening Trial (NLST) demonstrated a 20.0\% decrease in lung cancer specific mortality in the screen arm (annual LDCT) compared with the control arm (annual chest radiography) \citep{national2011reduced}. The trial enrolled over $50,000$ healthy subjects with more than 7 years of follow-up. In another example, the Heath Insurance Plan (HIP) mammography trial enrolled 62,000 women with 18 years of follow-up \citep{shapiro1997periodic}.

Surrogate (also known as alternative) endpoints, which can be determined at the time of cancer diagnosis (e.g. tumor stage or size), may facilitate faster evaluation of efficacy than mortality endpoints. Among possible surrogates, late-stage cancer incidence has emerged as perhaps the most promising choice \citep{dai2024strong,owens2022stage,feng2024cancer,dai2024clinical,neal2022cell,autier2009advanced}. Use of this surrogate is motivated by the fact that tumors diagnosed at early stage have better prognosis than tumors diagnosed at late stage. For example, ovarian cancer diagnosed at early-stage has approximately a 90\% 5-year survival probability while ovarian cancer diagnosed at a late-stage has approximately a 30\% 5-year survival probability \citep{WinNT}.

However, there is no guarantee that a screening modality which reduces late-stage incidence will also reduce mortality. For example, \cite{webb2024considerations} notes that certain biomarkers are only released by aggressive tumors and that these cancers, even if detected early, may remain deadly. A liquid biopsy screening test which detects these biomarkers may reduce late stage incidence but not affect mortality. In this case, late-stage reduction is a poor surrogate which could produce a misleading positive result.

This concern has prompted systematic investigation of the suitability of late-stage incidence as a surrogate for cancer-specific mortality, including meta-analyses of randomized screening trials \citep{dai2024strong,feng2024cancer,autier2009advanced}. These works have used a technique known as \textit{meta-analytic modeling} \citep{joffe2009related,vanderweele2013surrogate} to validate the late stage incidence surrogate. In these meta-analytic models, estimated mortality reduction was shown to correlate with estimated late-stage incidence reduction in  screening trials. Quantities such as regression slope and significance of positive association between primary and surrogate are reported. These regression analyses are often performed by cancer type (Ovarian, Breast, etc.), to evaluate possible cancer type specific differences in surrogate-primary correlations. For example \cite{feng2024cancer} found significant correlation between estimated mortality reduction and estimated late-stage incidence reduction in Ovarian screening trials (slope $1.16$, 95\% CI on Pearson correlation $(0.51,1.00)$) but not in colorectal screening trials (slope $0.4$, 95\% CI on Pearson correlation $(-0.27,0.80)$). Based on these results, \cite{feng2024cancer} concluded that late-stage incidence may be a suitable alternative endpoint for ovarian but not for colorectal cancer screening.

For each trial, the estimated mortality reduction and estimated late-stage incidence reduction generally do not exactly equal the true mortality reduction and true late-stage incidence reduction of the screening method, due to \textit{trial-level estimate uncertainty}. This uncertainty will be large in trials which enroll few subjects and/or with low cancer incidence. In this work, we demonstrate how trial-level estimate uncertainty affects inferences derived from meta-analytic models, with a focus on the practice of regressing estimated mortality reduction on estimated late-stage incidence reduction. We show that trial-level estimate uncertainty could bias regression slopes in a positive direction under the null hypothesis, potentially leading to a false indication of surrogacy. This positive bias is demonstrated in simulations and proved analytically. We then derive findings on the use of late-stage incidence surrogate in ovarian cancer, using confidence regions to assess the impact of uncertainty.

This work is organized as follows. In Section \ref{sec:meta-analysis} we review meta-analysis of surrogate and primary endpoints. In Section \ref{sec:notation}, we introduce notation. In Section \ref{sec:sim} we show via simulation that meta-analysis which ignores trial-level estimate uncertainty can lead to erroneous conclusions. In Section \ref{sec:analytic} we prove the generality of our simulation results under weak conditions. In Section \ref{sec:application} we construct 2-D confidence regions for the surrogate-primary pair and reassess the evidence for use of late-stage incidence as a surrogate endpoint in ovarian cancer. We conclude with a discussion in Section \ref{sec:discuss}.

\section{Meta-analytic Modeling of Surrogate Endpoints and trial-level Estimate Uncertainty}
\label{sec:meta-analysis}

Surrogate endpoints have been considered in clinical trials of treatments for many chronic diseases including AIDS, heart disease, and cancer \citep{prentice1989surrogate,daniels1997meta,gail2000meta,ellenberg1989surrogate,cuzick2007surrogate,etzioni2024revisiting,day1996trial,prasad2015strength}. Before use in practice, it is critical to evaluate the quality of a candidate surrogate endpoint and its relationship to the primary endpoint. \cite{joffe2009related} summarized four approaches to evaluating surrogate quality, including \textit{meta-analytic modeling} of completed studies. \cite{vanderweele2013surrogate} argued that among the four approaches, meta-analytic modeling may be the most promising.

Meta-analytic modeling of surrogates is complex and differs substantially from conventional meta-analysis of treatment efficacy. In meta-analysis of treatment efficacy, the objective is to pool information across trials to obtain an overall estimate of an efficacy parameter. For example, let $\widehat{M}_1, \ldots, \widehat{M}_{n_T}$ be estimated mortality reductions due to screening across $n_T$ trials. Under the assumption of \textit{homogeneity}, all screening methods share the same cancer mortality reduction $M$. The estimates $\widehat{M}_1, \ldots, \widehat{M}_{n_T}$ may be combined (e.g. using weighted average) to construct an estimate of $M$ which is more efficient than any individual estimate $\widehat{M}_l$. More sophisticated random effects models may be employed in the case of \textit{heterogeneity}, when the true efficacy $M$ varies across trials \citep{dersimonian1986meta,whitehead1991general}.

The purpose of meta-analytic modeling of surrogates is to estimate the functional relationship between a surrogate and primary endpoint across trials \citep{daniels1997meta,gail2000meta,burzykowski2005evaluation,joffe2009related,vanderweele2013surrogate}. Let $M_l$ denote the mortality reduction for trial $l$. For example if trial 1 is annual mammogram screening (with control arm receiving standard of care) while trial 2 is biannual mammograms (with control arm receiving standard of care), one may expect that $M_1 < M_2$, as more frequent mammograms are likely to induce greater mortality reduction. In addition to primary endpoints, each trial has surrogate endpoint $S_l$. In the cancer screening context, $S_l$ is late stage cancer incidence reduction. $S_l$ is also expected to vary across trials in the meta-analysis. In the mammography example, one may expect $S_1 < S_2$ because more frequent mammography will induce greater reduction in late-stage incidence. We refer to $M_l$ and $S_l$ as trial-level parameters because they are specific to trial $l$ and (while they can be estimated from trial data) are not known. The purpose of surrogate meta-analysis is to model the relationship between $M_l$ and $S_l$ across trials. For example, one may assume a linear relationship where
\begin{equation}
\label{eq:surrogate-model}
M_l =  \beta_0 + \beta_1 S_l + \epsilon_l
\end{equation}
and $\mathbb{E}[\epsilon_l] = 0$. As discussed in \cite{vanderweele2013surrogate}, heterogeneity in the surrogate endpoint ($S_l$) across trials is necessary for estimating $\beta_1$ and $\beta_0$ in Equation \eqref{eq:surrogate-model}, as regression estimates will be undefined if there is no variation in $S_l$. In the language of hierarchical modeling, $\beta_0$ and $\beta_1$ are hyperparameters because they are unknowns which model the distribution of trial specific parameters. Hyperparameters $\beta_0$ and $\beta_1$ can be clinically meaningful, because they indicate the amount of mortality reduction that can be expected from a given stage shift. According to \cite{vanderweele2013surrogate}, a good surrogate will explain a high proportion of variance in the primary (i.e. high $R^2$).

If trial level parameters $\{(S_l,M_l)\}_{l=1}^{n_T}$ were observed, one could estimate $\beta_0$ and $\beta_1$ with the ordinary least squares estimator:
\begin{equation}
\label{eq:oracle}
\widetilde{\beta}_0,\widetilde{\beta}_1 = \text{argmin}_{\beta_0,\beta_1} \sum_{l=1}^{n_T} \left(M_l - \beta_0 - \beta_1 S_l\right)^2.
\end{equation}
Statistical properties of $\widetilde{\beta}_0$ and $\widetilde{\beta}_1$, including consistency and asymptotic normality, are well known \citep{van2000asymptotic}. For example, under weak conditions $\widetilde{\beta}_1$ is an unbiased estimator of $\beta_1$.

However, the trial-level parameters $\{(S_l,M_l)\}_{l=1}^{n_T}$ are not known. Instead the estimated mortality reduction $\widehat{M}_l$ and estimated late stage incidence reduction $\widehat{S}_l$ in trial $l$ are observed. One thus cannot directly utilize the estimators in Equation \eqref{eq:oracle}. Note that $\widehat{M}_l \neq M_l$ and $\widehat{S}_l \neq S_l$ due to \textit{trial-level estimate uncertainty}. Statistical methodological exists for addressing this challenge. For example, hierarchical models have been proposed for estimating $\beta_0$ and $\beta_1$ in applications to therapeutics for heart disease and AIDS \citep{daniels1997meta,gail2000meta}.  These models do not require knowledge of parameters $M_l$ and $S_l$. However the models can be challenging to implement because they require patient level data for each trial, which may not be included in published trial results.
 
 Recent meta-analytic modeling in cancer screening \citep{dai2024strong,feng2024cancer} have directly regressed \textit{estimated} mortality reduction on \textit{estimated} late stage incidence reduction, ignoring the uncertainties of these estimated quantities. Formally,
\begin{equation}
\label{eq:practical}
\widehat{\beta}_0,\widehat{\beta}_1 = \text{argmin}_{\beta_0,\beta_1} \sum_{l=1}^{n_T} \left(\widehat{M}_l - \beta_0 - \beta_1 \widehat{S}_l\right)^2.
\end{equation}
These studies have used a standard t-test to test the null hypothesis of $H_0: \beta_1 = 0$ using $\widehat{\beta}_1$. Significance of the test has been presented as evidence that $S$ is a good candidate surrogate. However, the statistical properties of $\widehat{\beta}_0$ and $\widehat{\beta}_1$ as estimators of $\beta_0$ and $\beta_1$ are complex and have not been comprehensively studied. The following example illustrates the challenge.

\vspace{.1in}

\noindent
\underline{Null Example:} Suppose several screening trials have been conducted. In each trial, screening has no effect on the surrogate or primary endpoint: $S_l=M_l=0$ for $l=1,\ldots,n_T$. Estimators $\widetilde{\beta}_0$ and $\widetilde{\beta}_1$ are not uniquely defined in Equation \eqref{eq:oracle}. This indicates that the trials are non-informative as to the relationship between $M_l$ and $S_l$. \cite{vanderweele2013surrogate}'s stipulation that variation in $S_l$ is necessary for meta-analytic modeling is not satisfied. In practice, the problem is challenging because only estimates $\widehat{S}_l$ and $\widehat{M}_l$ are observed. None of these values are likely $0$. With $\widehat{S}_l$ and $\widehat{M}_l$, it is possible to use Equation \eqref{eq:practical} to compute $\widehat{\beta}_0$ and $\widehat{\beta}_1$. However the statistical properties of $\widehat{\beta}_0$ and $\widehat{\beta}_1$ are unknown and there are no (probabilistic) guarantees that one will reach the correct conclusion that these trials are uninformative as to the surrogate-primary relationship.



\section{Notation}
\label{sec:notation}

\begin{center}
\begin{figure}
 \centering
\includegraphics[width=0.95\linewidth]{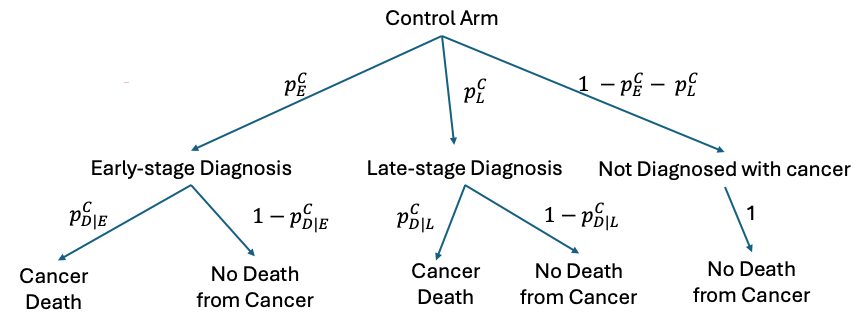}
\caption{Probability model for control arm. \label{fig:prob-model}}
\end{figure}
\end{center}
Consider a two-arm cancer screening trial with $n$ subjects in the control arm and $m$ subjects in the screen arm. For each subject, diagnosis (no cancer, early-stage, late-stage) and mortality (death from cancer, no death from cancer) outcomes are recorded. Figure \ref{fig:prob-model} displays parameters for the control arm. The probability of early-stage diagnosis in control is denoted as $p^C_E$ while the probability of late-stage diagnosis is $p^C_L$.  Define $p^C_{D|E}$ as probability of death due to cancer given an early-stage diagnosis in the control arm and $p^C_{D|L}$ as the probability of death due to cancer given a late-stage diagnosis in the control arm. In total, four parameters define outcomes in the control arm: $p^C_E,p^C_L,p^C_{D|E},p^C_{D|L}$. Analogous parameters apply to the screen arm of the trial with superscript $C$ replaced by $S$. For example, $p^S_{D|L}$ is the probability of death given late-stage diagnosis in the screen arm.

The surrogate measure of efficacy, reduction in late-stage incidence, is
\begin{equation}
\label{eq:sdef}
S = 1 - \frac{p^S_L}{p^C_L}.
\end{equation}
Cancer-specific mortality in the control and screen arms can be expressed as a function of the surrogate measure, respectively, 
\begin{align*}
    &p^C_D = p^C_Lp^C_{D|L} + p^C_Ep^C_{D|E}\\
    &p^S_D = p^S_Lp^S_{D|L} + p^S_Ep^S_{D|E}.
\end{align*}
The primary measure of efficacy, reduction in cancer-specific mortality, is
\begin{equation}
\label{eq:mdef}
    M = 1 - \frac{p^S_D}{p^C_D}.
\end{equation}
Let $X^C_i$ denote the stage (0 denoting no cancer diagnosis) at diagnosis for subject $i$ in the control arm with
\begin{equation*}
    X^C_{i} = \begin{cases} 
      0 & \text{no cancer diagnosis} \\
      1 & \text{early-stage diagnosis} \\
      2 & \text{late-stage diagnosis}. \\      
   \end{cases}
\end{equation*}
Let $Y^C_i$ denote the cancer mortality for subject $i$ in the control arm with
\begin{equation*}
    Y^C_{i} = \begin{cases} 
      0 & \text{no cancer death} \\
      1 & \text{cancer death}.      
   \end{cases}
\end{equation*}
Analogous definitions of $X^S_i$ and $Y^S_i$ apply to subject $i$ in the screen arm. Late-stage incidence reduction is estimated with
\begin{align*}
\widehat{p}^C_L &= \frac{1}{n} \sum_{i=1}^n 1\{X_i^C=2\}\\
\widehat{p}^S_L &= \frac{1}{m} \sum_{i=1}^m 1\{X_i^S=2\}\\
\widehat{S} &= 1 - \frac{\widehat{p}^S_L}{\widehat{p}^C_L}.
\end{align*}
Mortality reduction is estimated with
\begin{align*}
\widehat{p}^C_D &= \frac{1}{n} \sum_{i=1}^n 1\{Y_i^C=1\}\\
\widehat{p}^S_D &= \frac{1}{m} \sum_{i=1}^m 1\{Y_i^S=1\}\\
\widehat{M} &= 1 - \frac{\widehat{p}^S_D}{\widehat{p}^C_D}.
\end{align*}
For ease of notation, we defined these trial-level quantities (parameters, data, and estimators) without explicit dependence on trial.  When necessary, we use subscript $l$ to specify trial. For example, $\widehat{S}_l$ is the estimated late-stage incidence reduction in trial $l$ while $p^S_{D|E,l}$ is the probability of death given early stage diagnosis in the screen arm of trial $l$.

\section{Simulation}
\label{sec:sim}

Table \ref{tab:01-sim-params} displays four Scenarios for control arm and screen arm parameters ($p_E$, $p_L$, $p_{D|E}$, and $p_{D|L}$) in addition to the late stage incidence reduction ($S$) and mortality reduction ($M$) derived from these parameters. For example, in Scenario 1 all control arm and screen arm parameters are the same, thus there is no late stage incidence reduction and no mortality reduction so $S=M=0$.  For Scenario 3, the probability of early stage diagnosis in the control and screen arms are 0.010 and 0.013, respectively.

\begin{center}
    [Table 1 about here]
\end{center}

We conduct 4 simulations based on the Scenarios specified in Table \ref{tab:01-sim-params}. For each simulation, we sample $n_T=10$ trials and calculate $\widehat{\beta}_1$. We test the null hypothesis of $H_0: \beta_1=0$ using the t-statistic from the linear model fit (Equation \eqref{eq:practical}). We repeat this process $N=100$ times in each simulation. We compute $\mathbb{E}[\widehat{\beta}_1]$ (mean across $N$ repetitions) to assess bias and we compute the empirical probability of rejecting the null hypothesis $H_0: \beta_1=0$.

\begin{itemize}[label={},leftmargin=*]
\item \underline{Simulation A:} All $n_T$ trials are simulated from Scenario 1 with $n=m=20,000$ subjects per arm. Thus in every trial, screening has no effect on stage and no effect on mortality: $S_l=M_l=0$. 
\item \underline{Simulation B:} All $n_T$ trials are simulated from Scenario 2 with $n=m=20,000$ subjects per arm. Thus in every trial, screening reduces late stage incidence but has no effect on mortality: $S_l=0.13$ and $M_l=0$. 
\item \underline{Simulation C:} Each of the $n_T$ trials is simulated from one Scenario listed in Table \ref{tab:01-sim-params}, selected with equal probability. Thus each trial has one of 4 possible effects on late stage incidence and mortality. Each trial enrolls $n=m=20,000$ subjects per arm.
\item \underline{Simulation D:} Identical to Simulation C but with $n=m=100,000$ subjects per arm in each trial.
\end{itemize}

Results are shown in Table \ref{tab:01-sim-results}. Standard errors on Type I error rates are no larger than $\sqrt{0.5(1-0.5)/100}=0.05$. For Simulations A and B, the trials do not contain any information about the relationship between $M_l$ and $S_l$ because there is no variation in $S_l$ across trials. However $H_0: \beta_1=0$ is rejected in 94\% cases in Simulation A and 86\% of cases in Simulation B, giving a false conclusion that the trials indicate $S$ is a good surrogate. For Simulations C and D, there is substantial bias in the estimates of $\beta_1$. For Simulation C, this results in a substantially elevated Type I error rate of 17\%. Note that with the larger trial sample size in Simulation D, there is minimal inflation of Type I error.

\begin{center}
\begin{figure}
 \centering
   A) \includegraphics[width=0.45\linewidth]{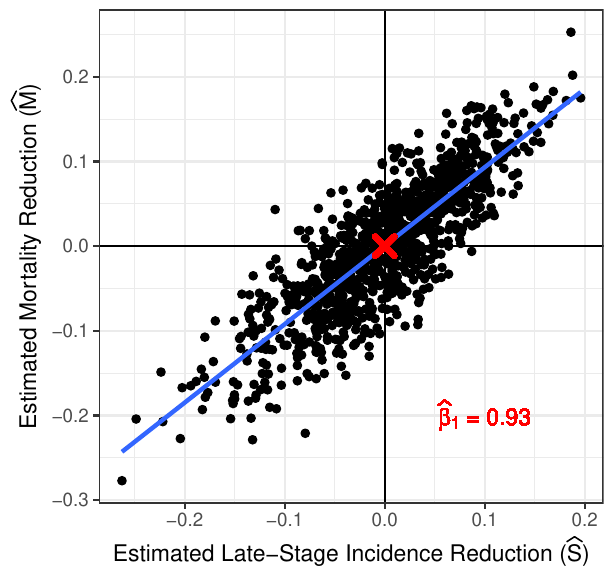} 
   B) \includegraphics[width=0.45\linewidth]{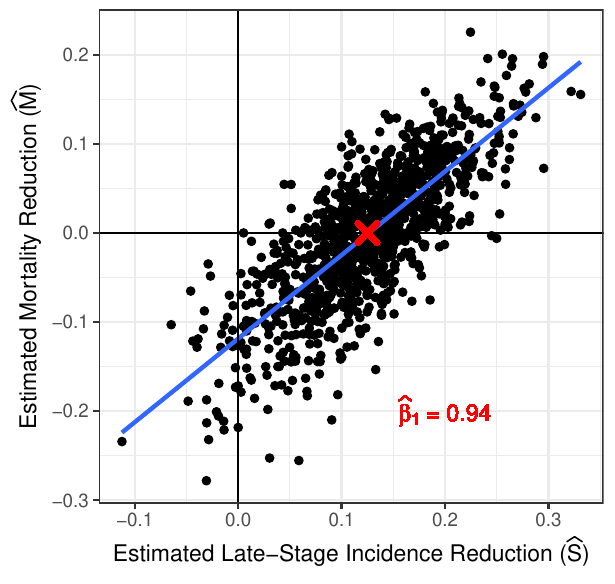}   \\
   C) \includegraphics[width=0.45\linewidth]{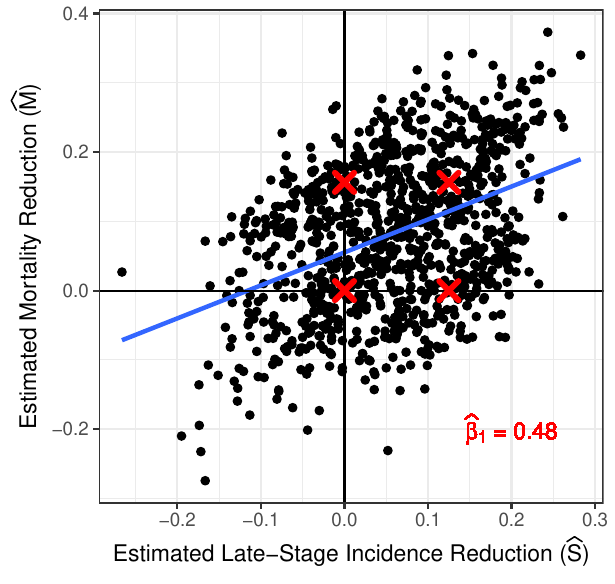}
   D) \includegraphics[width=0.45\linewidth]{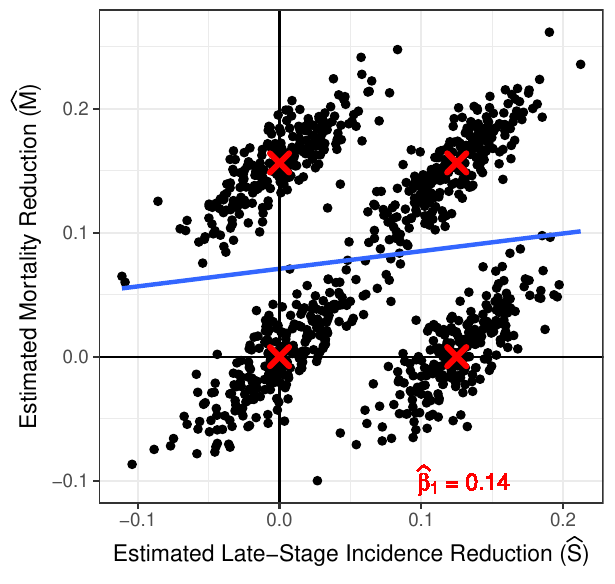}
\caption{Comparison of mortality reduction to reduction in late-stage incidence across four simulations. The black points are estimated mortality reduction and estimated late-stage incidence reduction. The red points are the parameter values. Regression lines (blue) and correlation ($R$) are computed by regressing estimated mortality reduction on estimated late stage incidence reduction. \label{fig:null-sim}}
\end{figure}
\end{center}

To better understand the results in Table \ref{tab:01-sim-results}, scatterplots of $\widehat{M}_l$ versus $\widehat{S}_l$ are displayed in Figure \ref{fig:null-sim} for the $n_T \times N$ trials generated in each Simulation. Trial parameters are displayed as red X. For example in Simulation A (Figure \ref{fig:null-sim}, panel A), $S_l=M_l=0$ for every trial so there is a single red X at $(0,0)$. 
The scatterplots show that the sampling distributions of $\widehat{S}_l$ and $\widehat{M}_l$ are positively correlated. For Simulations A and B, the estimator $\widehat{\beta}_1$ is completely determined by this sampling distribution because there is no variation in $S_l$ and $M_l$ across trials. 

\begin{center}
    [Table 2 about here]
\end{center}

For Simulations C and D the situation is more complex. Here both the sampling distribution of estimators $\widehat{S}_l$ and $\widehat{M}_l$ and genuine underlying variation in $S_l$ and $M_l$ across trials influence the estimator $\widehat{\beta}_1$. When trial sample sizes are larger (Simulation D), variation in $S_l$ and $M_l$ has a relatively stronger influence on $\widehat{\beta}_1$ than estimator sampling distributions, resulting in a less biased estimate and lower probability of rejecting the null.

Note that in these simulation results, weighting of regression estimators (weighted version of Equation \eqref{eq:practical}) by trial size, as has been proposed in the literature, will have no effect on the estimate $\widehat{\beta}_1$ as all trials within each simulation have an equal number of subjects (which implies that weighted and unweighted regression estimates are equivalent). Thus weighting is not a solution to the bias discussed here.

Note that for Simulations A and B, $\beta_1$ is not identifiable in Equation \eqref{eq:oracle} because there is no variation in $S_l$ (for Simulation A, $S_l=0$ and for Simulation B $S_l=0.13$). Simulation A is an instance of the Null Example proposed in Section \ref{sec:meta-analysis}, while Simulation B is a modification where there remains no variation in $S_l$. The correct conclusion for Simulation A is that the trials are non-informative as to the relationship between the primary and surrogate because no trial has had an effect on either endpoint. For Simulations C and D, $\beta_1=0$. The correct qualitative conclusion for Simulations C and D is that the meta-regression using the completed trials suggest $S$ is not a good surrogate of $M$. 

\section{Analytic Results}
\label{sec:analytic}

Biased slopes in the simulations in Section \ref{sec:sim} are caused by positive correlation in sampling distributions of estimators $\widehat{S}_l$ and $\widehat{M}_l$. We now show that this positive correlation holds under any plausible screening scenario. First note that $(\widehat{S}_l,\widehat{M}_l)$ is asymptotically normally distributed with mean of the trial parameters $(S_l,M_l)$.
\begin{theorem}
\label{thm:asympt}
\begin{equation*}
\sqrt{n}\left((\widehat{S}_l,\widehat{M}_l)^T - (S_l,M_l)^T\right) \rightarrow_d N(0,\Sigma_{SM,l})
\end{equation*}
\end{theorem}
See Appendix \ref{sec:reparam} and \ref{sec:asymp} for a proof. The asymptotic covariance is
\begin{equation}
\label{eq:smcov}
\Sigma_{SM,l} = 
\begin{pmatrix}
\Sigma_{S,l} & \rho_l\sqrt{\Sigma_{S,l}\Sigma_{M,l}} \\
\rho_l\sqrt{\Sigma_{S,l}\Sigma_{M,l}} & \Sigma_{M,l}
\end{pmatrix}
\end{equation}
where scalar parameters $\rho_l$, $\Sigma_{S,l}$, and $\Sigma_{M,l}$ are functions of trial-level parameters define in Section \ref{sec:notation}. The shape of the trial estimate estimate uncertainty (sampling distribution of $(\widehat{S}_l,\widehat{M}_l)$ is determined by $\rho_l$.
\begin{theorem}
\label{thm:cov}
Assume that in both control and screen arms, the probability of death given late-stage diagnosis is greater than the probability of death given early-stage diagnosis. Formally this assumption is $p^C_{D|E,l} < p^C_{D|L,l}$ and $p^S_{D|E,l} < p^S_{D|L,l}$. Then $\rho_l > 0$.
\end{theorem}
See Appendix \ref{sec:proof} for a proof. This result shows that the positive regression slopes observed in Simulations A and B (Figure \ref{fig:null-sim}) are not particular to the parameters selected in Scenarios 1 and 2 of Table \ref{tab:01-sim-params}, but rather a general property that holds under any plausible setting: the mortality rate for late-stage diagnoses is higher than that for early-stage diagnoses both control and screen arms. For more complex settings where there is variation in $S_l$ and $M_l$ across trials (e.g. Simulations C and D), the fact that $\rho_l > 0$ for each trial will bias estimates of $\beta_1$. The degree of this bias is determined by the variation in $S_l$ and $M_l$ across trials and the size of $\Sigma_{SM,l}$ for each trial.


\section{Meta-analytic Regression in Ovarian Screening Trials}
\label{sec:application}
To better assess how trial-level estimate uncertainty could affect conclusions in the meta-analytic regression, we propose visualizing uncertainty on estimates of primary and surrogate endpoints in screening trials. Construction of confidence sets for $(\widehat{S}_l,\widehat{M}_l)$ requires estimates of the three scalar parameters comprising $\Sigma_{SM,l}$ ($\Sigma_{S,l}$, $\Sigma_{M,l}$, and $\rho_l$) from Equation \eqref{eq:smcov}. These are functions of model parameters defined in Section \ref{sec:notation} and characterize the sampling distribution of the estimators for each trial. (See Appendix Section \ref{sec:cr} for mathematical details.) Diagonal elements ($\Sigma_{S,l}$ and $\Sigma_{M,l}$) are estimable from the marginal distribution of death and stage at diagnosis (by arm), information which is generally published in manuscripts and compiled for meta-analyses. However, $\rho_l$, which we term the sampling distribution correlation, depends on the conditional probability of death given stage of diagnosis in the screen and control arms of trial $l$. Estimates of these conditional probabilities is generally not provided in publications reporting the results of screening trials. Thus these conditional probabilities are not available for meta-analyses.

To address this limitation we conduct a sensitivity analysis, constructing confidence sets using various plausible values of the sampling distribution correlation $\rho_l$. As shown in Theorem \ref{thm:cov},  $\rho_l > 0$ under very general conditions. We select values $\rho_l = 0.1$, $0.66$, and $0.9$ for visualization with $0.1$ and $0.9$ representing extreme lower and upper bounds. The intermediate value of $0.66$ is used because in a subject level analysis of the National Lung Screening Trial (NLST), $\widehat{\rho}=0.66$ \citep{national2011reduced}. We plot results assuming all $n_T$ trials share the same $\rho_l$. In practice $\rho_l$ will vary across trials.

\begin{center}
\begin{figure}
 \centering
\includegraphics[width=1\linewidth]{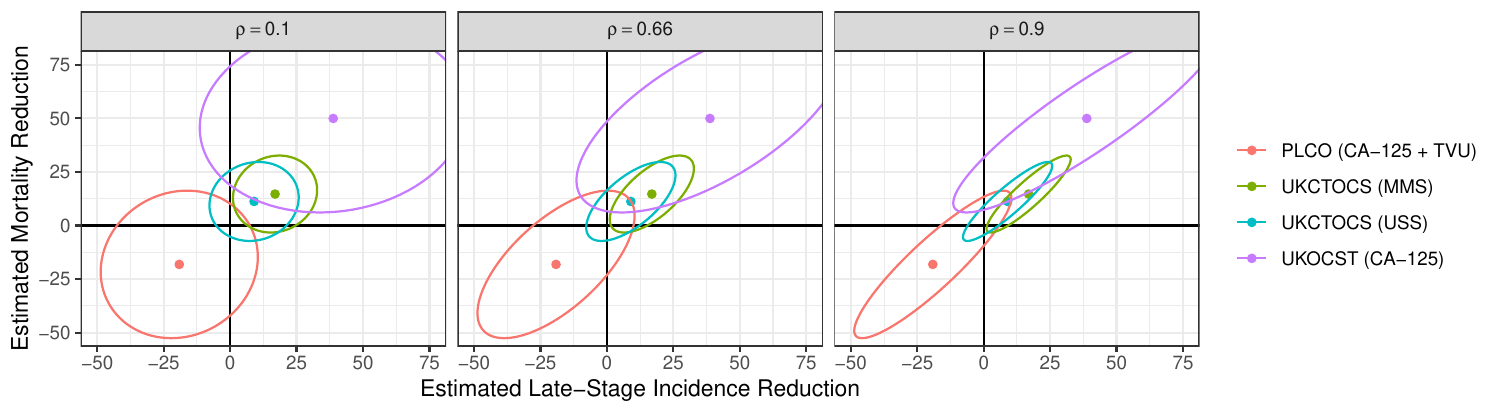}
\caption{Estimated mortality reduction versus estimated late-stage incidence reduction for ovarian screening trials with 90\% confidence sets for each trial estimates. Three values of $\rho_l$ are considered, assumed common across all four trials. \label{fig:ovarian}}
\end{figure}
\end{center}

\cite{feng2024cancer} regressed $\widehat{M}_l$ on $\widehat{S}_l$ in four screening trials for Ovarian cancer. Pearson correlation across four screening trials was estimated to be $R=0.99$ (95\% CI, 0.51 to 1.00) with linear regression slope of $\widehat{\beta}_1=1.15$ (Figure 1, Panel D from \cite{dai2024strong}). Figure \ref{fig:ovarian} displays a scatterplot using the estimated $(\widehat{S}_l,\widehat{M}_l)$ from the four ovarian cancer screening trials, adding 90\% confidence regions on the trial parameter estimates with $\rho_l=0.1$ (left plot), $\rho_l=0.66$ (center plot), and $\rho_l=0.9$ (right plot). PLCO and UKCTOCS (USS) confidence regions include $(0,0)$ at all correlations considered. UKCTOCS (MMS) confidence region contains $M=0$ at all correlations considered. The coverage properties of the UKOCST (CA-125) confidence region may not equal the nominal level of 90\% due to the small sample size of this trial (18 cancer deaths in control, 9 cancer deaths in screen). The size of the confidence regions shows that trial-level estimate uncertainty could lead to a false but significantly positive slope found in \cite{dai2024strong}. A seemingly positive association between trial-level late-stage incidence reduction (as a surrogate) and mortality reduction may not guarantee a useful surrogate for the mortality reduction, as demonstrated in the ovarian cancer screening trials.

\section{Discussion}
\label{sec:discuss}

It is well established from cancer registry databases that stage of cancer diagnosis and cancer lethality are highly correlated: cancers diagnosed at an early stage are more likely to be treated successfully, improving the overall prognosis. This fact has motivated research and development in cancer screening programs. This fact also motivates investigation of late stage incidence as a surrogate for reduction in mortality caused by screening. However, a strong correlation between two variables does not always imply that an intervention which alters one will alter the other. As an extreme counter-example, while ventricular arrhythemia and mortality are strongly correlated, drugs which have successfully reduced ventricular arrhythemia have been found to increase mortality \citep{moore1997deadly}. 

To avoid this so called ``surrogate paradox'' \citep{vanderweele2013surrogate}, meta-analytic modelings of cancer screening trials have sought to answer the questions: If a screening program reduces late stage incidence, will it also reduce mortality? If yes, by how much?  We showed via simulation and theory that trial-level estimate uncertainty could induce misleading positive slopes in meta-analytic regression models, even under the null setting of no reduction in both surrogate and primary outcomes across trials ($S_l=M_l=0$ for every trial). Existing work including \cite{feng2024cancer} has argued that the positive slopes in the meta-analysis regression provide affirmative answers to the aforementioned questions and thus support use of the proposed surrogate as an endpoint for future trials. This work calls for cautious interpretation of these claims.

Hierarchical models, such of those developed in \cite{daniels1997meta}, \cite{gail2000meta}, and \cite{buyse2000validation} could in principle be used to estimate regression parameters while accounting for trial-level estimate uncertainty. However there are several challenges to implementation in the context of cancer screening trials, including lack of access to subject level study data and limited number of studies in many cancer types.

This work was limited to studying the effect of trial-level estimate uncertainty on regression slopes in meta-analyses of cancer screening trials. We did not consider several additional challenges when evaluating suitability of the late-stage incidence surrogate. First, different functional forms between mortality reduction and late-stage incidence reduction may hold for different screening modalities. For example, colonoscopies and liquid biopsy screening for colorectal cancer may exhibit different relationships between primary and surrogate endpoints. There is no guarantee that either of these relationships will be linear or that there would even be a single linear relationship for all types of colonoscopy or all types of liquid biopsy. Given the limited number of trials in each cancer type and the uncertainty in the estimates of trial-level parameters, it is unlikely that these functional forms can be estimated with any reasonable degree of accuracy. Second, published results for cancer screening trials report subject outcomes with different follow-up times. Studies included in meta-analyses estimate endpoints at varying follow-up times. These differences in follow-up duration can alter the estimates of both primary and surrogate endpoints within each trial, potentially affecting the strength and consistency of correlations between primary and surrogate endpoints across trials.

Meta-analytic modeling has been used to study the quality of surrogates in many contexts, including therapeutics for cancer. \cite{prasad2015strength} reviewed 36 meta-analyses of surrogates across many therapeutics and disease sites. Each of these 36 meta-analyses performed meta-analytic regression and reported the strength of correlation between a primary (e.g. overall survival) and surrogate (e.g. progression free survival) outcome. Many of these studies did not incorporate trial-level estimate uncertainty, possibly influencing the strength of the estimated correlation between primary and surrogate endpoints. Adapting the simulations and theory developed here in Sections \ref{sec:sim} and \ref{sec:analytic} to survival outcomes may lead to a better understanding of the effect of trial-level uncertainty on meta-analysis outside of the cancer screening setting considered in this work.

For future studies, we recommend only using the meta-analysis regression to validate surrogates when trial-level uncertainty can be propagated into model fits, as done in methodology developed by \cite{daniels1997meta} and \cite{gail2000meta}. When this is not possible, alternative assessments of surrogate quality may be useful, such as methods proposed in \cite{joffe2009related}. When performed, meta-analytic modeling should include plots with size of trial-level estimate uncertainty as we have done in Figure \ref{fig:ovarian}.

\section{Appendix}
\label{sec:appendix}

\subsection{Reparameterization of Model}
\label{sec:reparam}

In Section \ref{sec:notation}, the model is parameterized using conditional probabilities, e.g. $p^C_{D|E}$ is the probability of death given early-stage diagnosis in the control group. For the purpose of deriving asymptotic results, it is simpler to parameterize the model in terms of joint probabilities of occurrence of both primary mortality endpoint (death from cancer, no death from cancer) and surrogate endpoint (no cancer diagnosis, early-stage diagnosis, late-stage diagnosis).

There are a total of 6 events (2 mortality outcomes $\times$ 3 diagnosis outcomes) per arm. However, since it is not possible for an individual to die of cancer without being diagnosed with cancer, only 5 of these events have non-zero probability. Let $p^C_{jk}$ be the probability of mortality endpoint $j$ ($j=0$ is no death from cancer, $j=1$ is death from cancer) and diagnosis endpoint $k$ ($k=0$ is no cancer diagnosis, $k=1$ is early-stage diagnosis, and $k=2$ is late-stage diagnosis) in the control arm. Identical definitions apply to the screen arm with superscript $C$ replaced by $S$.

One can map the conditional parameterization in Section \ref{sec:notation} to the joint parameterization with 

\begin{align*}
p^C_{00} &= 1-p^C_E-p^C_L\\
p^C_{10} &= 0\\
p^C_{01} &= p^C_E(1-p^C_{D|E})\\
p^C_{11} &= p^C_E p^C_{D|E}\\
p^C_{02} &= p^C_L(1-p^C_{D|L})\\
p^C_{12} &= p^C_L p^C_{D|L}
\end{align*}
Estimators are
\begin{align*}
\widehat{p}^C_{jk} = \frac{1}{n}\sum 1_{X^C_i=j,Y^C_i=k}\\ 
\widehat{p}^S_{jk} = \frac{1}{n}\sum 1_{X^S_i=j,Y^S_i=k}.
\end{align*}

\subsection{Proof of Theorem 1}
\label{sec:asymp}

Define $p=(p_{11}^C,p_{02}^C,p_{12}^C,p_{11}^S,p_{02}^S,p_{12}^S)^T$ and $\widehat{p} = (\widehat{p}_{11}^C,\widehat{p}_{02}^C,\widehat{p}_{12}^C,\widehat{p}_{11}^S,\widehat{p}_{02}^S,\widehat{p}_{12}^S)^T$. By the Central Limit Theorem
\begin{equation*}
\sqrt{n}(\widehat{p} - p) \rightarrow_d N(0,\Sigma)
\end{equation*}
where
\begin{align}
\label{eq:fullcov}
\Sigma_C &= \begin{pmatrix}
p^C_{11}(1-p^C_{11}) & - p^C_{11}p^C_{02} & - p^C_{11}p^C_{12}\\
& p^C_{02}(1-p^C_{02}) & - p^C_{02}p^C_{12}\\
& & p^C_{12}(1-p^C_{12})
\end{pmatrix}\\ \nonumber
\Sigma_S &= \frac{n}{m}\begin{pmatrix}
p^S_{11}(1-p^S_{11}) & - p^S_{11}p^S_{02} & - p^S_{11}p^S_{12}\\
& p^S_{02}(1-p^S_{02}) & - p^S_{02}p^S_{12}\\
& & p^S_{12}(1-p^S_{12})
\end{pmatrix}\\ \nonumber
\Sigma &= \begin{pmatrix}
\Sigma_C & 0 \\
0 & \Sigma_S
\end{pmatrix}.
\end{align}
The parameters of interest (late-stage reduction and mortality reduction) are
\begin{equation*}
g(p) = (g_1(p),g_2(p))^T = (S,M)^T = \left(1 - \frac{p^S_{02} + p^S_{12}}{p^C_{02} + p^C_{12}}, 1-\frac{p^S_{11} + p^S_{12}}{p^C_{11} + p^C_{12}} \right)^T.
\end{equation*}
By application of the Delta Method,
\begin{equation*}
\sqrt{n}\left((\widehat{S},\widehat{M})^T - (S,M)^T\right) \rightarrow_d N(0,\Sigma_{SM})
\end{equation*}
where
\begin{equation}
\label{eq:smcov2}
\Sigma_{SM} = \nabla g(p)^T\Sigma \nabla g(p).
\end{equation}

\subsection{Proof of Theorem 2}
\label{sec:proof}

We show $Cov(\widehat{S},\widehat{M}) = \Sigma_{SM,12} > 0$. From Equation \eqref{eq:smcov2} we have
\begin{equation*}
\Sigma_{SM,12} = \nabla g_1(p)^T\Sigma \nabla g_2(p).
\end{equation*}
Note that
\begin{equation*}
\nabla g_1(p) = \begin{pmatrix}
0 \\
\frac{-(p^S_{02} + p^S_{12})}{(p^C_{02}+p^C_{12})^2}\\
\frac{-(p^S_{02} + p^S_{12})}{(p^C_{02}+p^C_{12})^2}\\
0\\
\frac{1}{p^C_{02} + p^C_{12}}\\
\frac{1}{p^C_{02} + p^C_{12}}
\end{pmatrix} = 
\begin{pmatrix}
0 \\
r\\
r\\
0\\
s\\
s
\end{pmatrix} \text{\, \, \, and \, \, \,}
\nabla g_2(p) = \begin{pmatrix}
\frac{-(p^S_{11} + p^S_{12})}{(p^C_{11} + p^C_{12})^2} \\
0\\
\frac{-(p^S_{11} + p^S_{12})}{(p^C_{11} + p^C_{12})^2} \\
\frac{1}{p^C_{11} + p^C_{12}}\\
0\\
\frac{1}{p^C_{11} + p^C_{12}}
\end{pmatrix} = 
\begin{pmatrix}
t \\
0\\
t\\
u\\
0\\
u
\end{pmatrix}.
\end{equation*}
Using these derivations and the asymptotic covariance of $\widehat{p}$ derived in Equation \eqref{eq:fullcov} we have
\begin{align*}
\Sigma_{SM,12} &= \underbrace{(-p^C_{11}p^C_{02} - p^C_{11}p^C_{12} - p^C_{02}p^C_{12} + p^C_{12}(1-p^C_{12}))}_{\equiv A}rt \\
&+ \underbrace{(-p^S_{11}p^S_{02} - p^S_{11}p^S_{12} - p^S_{02}p^S_{12} + p^S_{12}(1-p^S_{12}))}_{\equiv B}su(n/m).
\end{align*}
Note that $r,t < 0$ and $s,u > 0$. Thus $rt > 0$ and $su > 0$. Thus to prove $\Sigma_{SM,12} > 0$, it is sufficient to show that $A,B > 0$. Since $A$ and $B$ are defined analogously for control and screen, sufficient to show $A > 0$.

Note that $p^C_{D|E} = p^C_{11} / (p^C_{11} + p^C_{01})$ and $p^C_{D|L} = p^C_{12} / (p^C_{12} + p^C_{02})$. Thus the theorem assumption is equivalent to 
\begin{equation*}
\frac{p^C_{11}}{p^C_{11} + p^C_{01}} < \frac{p^C_{12}}{p^C_{12} + p^C_{02}}
\end{equation*}
which implies
\begin{equation}
\label{eq:temp1}
\frac{p^C_{11}}{p^C_{11} + p^C_{01} + p^C_{00}} < \frac{p^C_{12}}{p^C_{12} + p^C_{02}}.
\end{equation}

\begin{align*}
A &= -p^C_{11}p^C_{02} - p^C_{11}p^C_{12} - p^C_{02}p^C_{12} + p^C_{12}(p^C_{00} + p^C_{01} + p^C_{11} + p^C_{02})\\
&= p^C_{12}p^C_{01} + p^C_{12}p^C_{00} - p^C_{11}p^C_{02}\\
&>0
\end{align*}
where the final inequality follows from Inequality \ref{eq:temp1}.

\subsection{Confidence Region for $(S,M)$}
\label{sec:cr}
A $(1-\alpha)100$\% Wald Confidence Region for $(S,M)$ is
\begin{equation*}
C_{SM}^\alpha = \left\{(S,M) : \left((S,M) - (\widehat{S},\widehat{M})\right) \widehat{\Sigma}_{SM}^{-1}  \left((S,M) - (\widehat{S},\widehat{M})\right)^T < q_2^{\alpha}\right\}
\end{equation*}
where $q_2^\alpha$ is the $(1-\alpha)$ quantile of a chi-squared distribution with $2$ degrees of freedom and $\widehat{\Sigma}_{SM}$ is a consistent estimator of $\Sigma_{SM}$ (defined in Equation \eqref{eq:smcov2}). Note that $\Sigma_{SM}$ consists of three scalar quantities ($\Var(\widehat{S})$, $\Var(\widehat{M})$, and $\rho$). The quantity $\rho$ cannot be estimated from data available in meta-analysis. However $\Var(\widehat{S})$ and $\Var(\widehat{M})$ can be estimated. Specifically, recall that
\begin{equation*}
M = 1-\frac{p^S_D}{p^C_D}
\end{equation*}
where  $p^S_D$ is the probability of mortality in the screen arm and $p^C_D$ is the probability of mortality in the control arm. By the CLT
\begin{equation*}
\sqrt{n}\left((\widehat{p}^C_D,\widehat{p}^S_D)^T - (p^C_D,p^S_D)^T\right) \rightarrow_d N\left(0,
\begin{pmatrix}
p^C_D(1-p^C_D) & 0 \\
0 & (n/m)p^S_D(1-p^S_D)
\end{pmatrix}
\right)
\end{equation*}
By application of the delta method with $M = f(p_D^C,p_D^S) = 1 - p_D^S/p_D^C$,
\begin{equation*}
\sqrt{n}(\widehat{M} - M) \rightarrow_d N\left(0,\left(\frac{p^S_D}{p^C_D}\right)^2p_D^C(1-p_D^C) + \frac{n}{m}\frac{1}{(p_D^C)^2}p_D^S(1-p_D^S)\right).
\end{equation*}
By application of the delta method with $S = f(p_S^C,p_S^S) = 1 - p_S^S/p_S^C$,
\begin{equation*}
\sqrt{n}(\widehat{S} - S) \rightarrow_d N\left(0,\left(\frac{p^S_S}{p^C_S}\right)^2p_S^C(1-p_S^C) + \frac{n}{m}\frac{1}{(p_S^C)^2}p_S^S(1-p_S^S)\right).
\end{equation*}



\section*{Acknowledgements, Data Rights, and Reproducibility}

JPL received support from the National Cancer Institute and the National Center for Advancing Translational Sciences of the NIH [P30CA016672 and CCTS UM1TR004906], YS is partially supported by NCI BRG (P30CA016672). NLST data was accessed under project NLST-2696 \url{https://cdas.cancer.gov/approved-projects/2696/}.

Code for reproducing the results in the work is available \url{https://github.com/longjp/surrogate-code}.

\bibliographystyle{rss}
\bibliography{refs}

\clearpage

\begin{table}

\caption{\label{tab:unnamed-chunk-24}Trial parameters.  \label{tab:01-sim-params}}
\centering
\begin{tabular}[t]{llrrrrrr}
\toprule
 & Arm & $p_E$ & $p_L$ & $p_{D|E}$ & $p_{D|L}$ & S & M\\
\midrule
Scenario 1 & Control & 0.010 & 0.020 & 0.10 & 0.750 & 0.000 & 0.000\\
 & Screen & 0.010 & 0.020 & 0.10 & 0.750 &  & \\
Scenario 2 & Control & 0.010 & 0.020 & 0.10 & 0.750 & 0.125 & 0.000\\
 & Screen & 0.012 & 0.017 & 0.28 & 0.714 &  & \\
Scenario 3 & Control & 0.010 & 0.020 & 0.10 & 0.750 & 0.125 & 0.156\\
 & Screen & 0.013 & 0.017 & 0.08 & 0.714 &  & \\
Scenario 4 & Control & 0.010 & 0.020 & 0.10 & 0.750 & 0.000 & 0.156\\
 & Screen & 0.010 & 0.020 & 0.10 & 0.625 &  & \\
\bottomrule
\end{tabular}
\end{table}

\clearpage

\begin{table}

\caption{\label{tab:unnamed-chunk-25}Simulation results.  \label{tab:01-sim-results}}
\centering
\begin{tabular}[t]{llrr}
\toprule
  & $\beta_1$ & $\mathbb{E}[\widehat{\beta}_1]$ & Type I Error\\
\midrule
Simulation A & - & 0.93 & 0.94\\
Simulation B & - & 0.96 & 0.86\\
Simulation C & 0 & 0.46 & 0.17\\
Simulation D & 0 & 0.12 & 0.07\\
\bottomrule
\end{tabular}
\end{table}

\end{document}